\documentstyle[12pt]{article}

\begin{document}
{\sf \begin{center} \noindent
{\Large \bf Suppressed fluctuations in non-stretched-twist-fold turbulent helical dynamos}\\[3mm]

by \\[0.3cm]

{\sl L.C. Garcia de Andrade}\\

\vspace{0.5cm} Departamento de F\'{\i}sica
Te\'orica -- IF -- Universidade do Estado do Rio de Janeiro-UERJ\\[-3mm]
Rua S\~ao Francisco Xavier, 524\\[-3mm]
Cep 20550-003, Maracan\~a, Rio de Janeiro, RJ, Brasil\\[-3mm]
Electronic mail address: garcia@dft.if.uerj.br\\[-3mm]
\vspace{2cm} {\bf Abstract}
\end{center}
\paragraph*{}
Suppression of fluctuations of normally perturbed magnetic fields in
dynamo waves and slow dynamos along curved (folded), torsioned
(twisted) and non-stretched, diffusive filaments are obtained. This
form of fluctuations suppression has been recently obtained by
Vainshtein et al [PRE 56, (1997)] in nonlinear ABC and
stretch-twist-fold (STF) dynamos by using a magnetic Reynolds number
of the order of $Rm\approx{10^{4}}$. Here when torsion does not
vanish an expression between magnetic Reynolds number and length
scale L as with constant torsion ${\tau}_{0}$ itself is obtained,
such as $Rm\approx{\frac{{\tau}_{0}L}{\eta}}$ is obtained. At
coronal loops $Rm\approx{10^{12}}$ and torsion of the twisted
structured loop from astronomical data by Lopez-Fuentes et al
[Astron. and Astrophys. (2003)] of
${\tau}\approx{9.0{\times}10^{-10}}cm^{-1}$ is used to compute a
very slow magnetic diffusion of ${\eta}\approx{10^{-8}}$. The slow
dynamo obtained here is in agreement with Vishik arguement that fast
dynamo cannot be obtained in non-stretched dynamo flows. When
torsion vanishes helical turbulence is quenched and but
${\alpha}$-dynamos cannot be maintained since exponential stretching
depends on torsion. This is actually Zeldovich antidynamo theorem
for torsion-free or planar filaments which has been discussed by the
other also recently in another context [Astr Nach. (2008)]. The
suppression of magnetic field fluctuations is actually a result of
the coupling of the magnetic diffusion and Frenet torsion of helical
turbulence.{\bf PACS
numbers:\hfill\parbox[t]{13.5cm}{02.40.Hw:differential geometries.
91.25.Cw-dynamo theories.}}

\newpage
\newpage
 \section{Introduction}
 Earlier M. Vishik \cite{1} has argued that only slow dynamos can be obtained from non-stretching dynamo flows and no fast dynamos so
 well-known to be obtained from the  Vainshtein and Zeldovich \cite{2} work on stretch-twist and fold (STF) \cite{3}
 magnetic dynamo generation mechanism on ropes and
 magnetic filaments. Ropes are essentially thin twisted tubes filled with plasma flow and or magnetic
 fields which are compressed by the stretching of the tube
 giving rise to an amplification of the magnetic field. Diffusion processes
 which have been investigated in the context of Riemannian geometry by S. Molchanov \cite{4} have also been given strongly
 importance in obtaining slow dynamos. Such slow dynamos which have previously obtained by Soward \cite{5} by simply
 aligning magnetic and flow velocity to obtain helicity distinct from zero. In this report one finds two solutions to the self-induction equations
 , first an analytical solution which is a torsioned turbulent filament with very small torsion and diffusion. When torsion vanishes
 filaments are planar and by Zeldovich anti-dynamo theorem \cite{6} cannot support dynamo action, actually it is shown that it
 generates a static magnetic initial field and a steady perturbation which may be a marginal dynamo at maximum.
 Thus as the same way fast dynamo are generated by stretch, folding
 and twisting of the loops or filaments it seems that non-stretched
 , fold and twisted filaments leads to slow dynamos filaments. The paper is organized as
 follows: In section 2 a brief review on dynamics of holonomic Frenet frame is presented. In section
 3 the self-induction equation is solved in this frame for slow dynamos and magnetic torsion is obtained from Rm
 and solutions are shown in the torsion-free case to be consistent with Zeldovich anti-dynamo theorem. In this
 section one also shows that chaos is restricted \cite{7} in the ${\alpha}-dynamos$. In section 4
 conclusions are presented.
\newpage
\section{Non-stretched dynamo filamentary flows in Frenet frame} This section deals with a very brief review of the Serret-Frenet
holonomic frame \cite{8} equations that are specially useful in the
investigation of STF Riemannian flux tubes in magnetohydrodynamics
(MHD) with magnetic diffusion. Frenet frame has been used by solar
physicists and astrophysicists \cite{8,9}to investigate twisted
solar flux tubes in the solar photosphere. Here the Frenet frame is
attached along the magnetic flux tube axis which possesses Frenet
torsion and curvature \cite{7}, which completely determine
topologically the filaments, one needs some dynamical relations from
vector analysis and differential geometry of curves such as the
Frenet frame $(\textbf{t},\textbf{n},\textbf{b})$ equations
\begin{equation}
\textbf{t}'=\kappa\textbf{n} \label{1}
\end{equation}
\begin{equation}
\textbf{n}'=-\kappa\textbf{t}+ {\tau}\textbf{b} \label{2}
\end{equation}
\begin{equation}
\textbf{b}'=-{\tau}\textbf{n} \label{3}
\end{equation}
The holonomic dynamical relations from vector analysis and
differential geometry of curves by
$(\textbf{t},\textbf{n},\textbf{b})$ equations in terms of time
\begin{equation}
\dot{\textbf{t}}=[{\kappa}'\textbf{b}-{\kappa}{\tau}\textbf{n}]
\label{4}
\end{equation}
\begin{equation}
\dot{\textbf{n}}={\kappa}\tau\textbf{t} \label{5}
\end{equation}
\begin{equation}
\dot{\textbf{b}}=-{\kappa}' \textbf{t} \label{6}
\end{equation}
along with the flow derivative
\begin{equation}
\dot{\textbf{t}}={\partial}_{t}\textbf{t}+(\vec{v}.{\nabla})\textbf{t}
\label{7}
\end{equation}
From these equations and the generic flow
\begin{equation}
\dot{\textbf{X}}=v_{s}\textbf{t}+v_{n}\textbf{n}+v_{b}\textbf{b}
\label{8}
\end{equation}
one obtains
\begin{equation}
\frac{{\partial}l}{{\partial}t}=(-\kappa{v}_{n}+{v_{s}}')l\label{9}
\end{equation}
where l is given by
\begin{equation}
l:=(\textbf{X}'.\textbf{X}')^{\frac{1}{2}}\label{10}
\end{equation}
which shows that if $v_{s}$ is constant, which fulfills the
solenoidal incompressible flow
\begin{equation}
{\nabla}.\textbf{v}=0\label{11}
\end{equation}
and $v_{n}$ vanishes, one should have an non-stretched twisted flux
tube. This is exactly the choice $\textbf{v}=v_{0}\textbf{t}$, where
$v_{0}=constant$ is the steady flow one uses here. The solution
\begin{equation}
\textbf{B}=B_{s}(s,t)\textbf{t}\label{12}
\end{equation}
shall be considered here. This definition of magnetic filaments is
shows from the solenoidal carachter of the magnetic field
\begin{equation}
{\nabla}.\textbf{B}=0\label{13}
\end{equation}
where $B_{s}$ is the toroidal component of the magnetic field. In
the next section one shall solve the diffusion equation in the
steady case in the non-holonomic Frenet frame as
\begin{equation}
{\partial}_{t}\textbf{B}={\nabla}{\times}(\textbf{v}{\times}\textbf{B})+{\eta}{\nabla}^{2}\textbf{B}
\label{14}
\end{equation}
where ${\eta}$ is the magnetic diffusion. Since in astrophysical
scales ,
${\eta}{\nabla}^{2}\approx{{\eta}{L^{-2}}}\approx{{\eta}{\times}10^{-20}}cm^{-2}$
for a solar loop scale length of $10^{10}cm$ \cite{6} one notes that
the diffusion effects are not highly appreciated in astrophysical
dynamos, though they are not neglected here. Let us now consider the
magnetic field definition in terms of the magnetic vector potential
$\textbf{A}$ as
\begin{equation}
\textbf{B}={\nabla}{\times}\textbf{A} \label{15}
\end{equation}
the gradient operator is
\begin{equation}
{\nabla}=\textbf{t}{\partial}_{s}\label{16}
\end{equation}
\newpage
\section{Turbulent helical dynamos and dynamo waves in filaments}
In this section we review the magnetic perturbation which leads to
the dynamo waves in filaments \cite{10}. Let us start by assuming
that the filaments are perturbed accordingly to the laws
\begin{equation} \textbf{B}={\textbf{B}}_{0}+{\textbf{B}}_{1} \label{17}
\end{equation}
where $\textbf{B}_{0}=B_{0}\textbf{t}$ is the initial field and
$\textbf{B}_{1}$ is the normally perturbed magnetic field w.r.t to
the filament axis itself. This field in Frenet frame is given by
\begin{equation}
\textbf{B}_{1}=b_{1}\textbf{n}+b_{2}\textbf{b} \label{18}
\end{equation}
By performing the substitution into the turbulent structure of the
magnetic induction equation in the steady state
\begin{equation}
{\eta}{\nabla}^{2}\textbf{B}_{1}+(\textbf{B}_{0}.{\nabla})\textbf{v}=0
\label{19}
\end{equation}
where one has considered that
$\textbf{V}=\textbf{V}_{0}+\textbf{v}$, and  $\textbf{V}_{0}$
vanishes. The turbulent averages $<\textbf{B}_{1}>$ and
$<\textbf{v}>$ both vanish and term \begin{equation}
\textbf{v}{\times}\textbf{B}_{1}-<\textbf{v}{\times}\textbf{B}_{1}>=0
\label{20}
\end{equation}
By taking Parker's ${\alpha}$-effect hypothesis
\begin{equation}
<\textbf{v}{\times}\textbf{B}_{1}>={\alpha}\textbf{B}_{0} \label{21}
\end{equation}
the remaining equation is
\begin{equation}
{\partial}_{t}\textbf{B}_{0}={\eta}{\nabla}^{2}\textbf{B}_{0}+rot({\alpha}\textbf{B}_{0})
\label{22}
\end{equation}
which is the ${\alpha}$-dynamo equation. Here the term
$rot({\textbf{V}_{0}{\times}\textbf{B}_{0}})$ vanishes since
$\textbf{V}_{0}$. The first equation one shall solve is
$div\textbf{B}=0$ which yields
\begin{equation}
{\partial}_{s}{B}_{0}={\tau}_{0}b_{1} \label{23}
\end{equation}
where torsion ${\tau}_{0}$ and curvature ${\kappa}_{0}$ of helical
filament coincides and are constants. \newpage Taking the dynamo
ansatz $\textbf{B}_{0}=\textbf{b}_{0}e^{{\omega}t+ik_{s}s}$ , where
the constant factor $k_{s}$ is the dynamo wave number, and
${\omega}$ plays the role of dynamo factor, we may solve the
equation (\ref{22}). By examining this factor below in the limit of
$Rm\rightarrow{\infty}$, one shall be able to check if the dynamo is
fast or slow. Equation (\ref{22}) now yields
\begin{equation}
{{\partial}_{s}}^{2}\textbf{B}_{0}=-{k_{s}}^{2}\textbf{B}_{0}
\label{24}
\end{equation}
and
\begin{equation}
B_{0}{\tau}_{0}(v_{0}-{\tau}_{0})\textbf{n}+\textbf{B}_{0}{\omega}=\textbf{B}_{0}{\alpha}{\tau}_{0}{\times}\textbf{n}
- i{k^{2}}_{s}{\alpha}]-{\eta}{k_{s}}^{2}\textbf{B}_{0}\label{25}
\end{equation}
Using the equations above for the dynamics of Frenet frame and
transvecting this equation with scalar product of $\textbf{B}_{0}$
yields
\begin{equation}
{\omega}=-{\eta}{k_{s}}^{2}\label{26}
\end{equation}
which is similar to Parker's dynamo wave frequency , with only basic
difference that now it is a real number. The phase velocity is
\begin{equation}
v_{ph}:=\frac{{\omega}}{k_{s}}=-{\eta}{k_{s}}\label{27}
\end{equation}
This is already an indication that the dynamo is slow since it
violates the condition for fast dynamo cause this condition is given
by the fact that ${\omega}>{0}$ in the limit of
$Rm\rightarrow{\infty}$, which is certainly not the case and
${\omega}\le{0}$ since the diffusivity $\eta$ is always positive.
Since $Rm=\frac{v_{0}L}{\eta}$ actually ${\omega}$ vanishes and the
dynamo is slow or marginal in this case. Nevertheless since
diffusivity here is finite and $Rm$ is also finite a non-trivial
solution may be obtained here. Note that by transvecting the
equation (\ref{22}) with the vector product of $\textbf{B}_{0}$ one
obtains
\begin{equation}
[\textbf{B}_{0}{\alpha}{\tau}_{0}{\times}\textbf{n}]{\times}\textbf{B}_{0}=0\label{28}
\end{equation}
This expression is fundamental since implies that
${\tau}_{0}{\alpha}=0$, which means that either ${\alpha}$ effect
vanishes or torsion vanishes. We rather choose the first option
since we shall show that torsion suppresses dynamo effect
fluctuations. Transvecting equation (\ref{22}) again with
$\textbf{n}$ yields that
\begin{equation}
B_{0}{\tau}_{0}(v_{0}-{\tau}_{0})= 0\label{29}
\end{equation}
where we have used simple vector expressions
$\textbf{B}_{0}.\textbf{n}=0$ and
$\textbf{B}_{0}{\times}\textbf{B}_{0}=0$.
\newpage
Expression (\ref{29}) then yields ${\tau}_{0}=v_{0}$. The remaining
equation for $\textbf{B}_{1}$ yields
\begin{equation}
{\eta}[{b"}_{1}+2{\tau}_{0}({b'}_{1}-{b'}_{2})-b_{2}{{\tau}_{0}}^{2}]+v_{0}{\tau}_{0}B_{0}=0\label{30}
\end{equation}
\begin{equation}
{b"}_{2}-b_{2}{{\tau}_{0}}^{2}=0\label{31}
\end{equation}
\begin{equation}
2{b'}_{1}-b_{2}{{\tau}_{0}}=0\label{32}
\end{equation}
Summing up the last two equations yields
\begin{equation}
{b"}_{1}=2{b'}_{1}{{\tau}_{0}}\label{33}
\end{equation}
Equation (\ref{24}) yields
\begin{equation}
{b'}_{1}=i\frac{k_{s}}{{\tau}_{0}}B_{0}\label{34}
\end{equation}
Integration of this expression yields $b_{1}(s,t)$ as
\begin{equation}
{b}_{1}=\frac{1}{{\tau}_{0}}B_{0}\label{35}
\end{equation}
Substitution of (\ref{35}) and (\ref{35}) into equation (\ref{30})
yields
\begin{equation}
i[{k^{4}}_{s}+2{{\tau}_{0}}k_{s}]+2[{k^{3}}_{s}+5{{\tau}_{0}}^{2}k_{s}]=0\label{36}
\end{equation}
where one has assumed that the wave length $k_{s}$ can be a complex
number. In the limit of long wave-length , the dynamo wave number
modulus is $|k_{s}|<<1$, and we approximate
$O({k^{4}}_{s})=O({k^{3}}_{s})\approx{0}$, this result yields the
equation
\begin{equation}
i2{{\tau}_{0}}{k^{2}}_{s}+5{{\tau}_{0}}^{2}k_{s}=0\label{37}
\end{equation}
which yields
\begin{equation}
{k_{s}}= i\frac{5}{2}{{\tau}_{0}}\label{38}
\end{equation}
which shows that the exponential stretching of the magnetic field
$\textbf{B}_{1}$ is strongly suppressed since
\begin{equation}
{b}_{1}=\frac{1}{{\tau}_{0}}e^{-[\frac{5}{2}[{\tau}_{0}(s-{\eta}{\tau}_{0}t)]]}\label{39}
\end{equation}
note that since torsion of the solar loop is rather small, the
second term in the exponent of this expression takes a long time to
increase since ${\eta}$ is also rather weak. Actually the formula
(\ref{39}) shows that the fluctuation $b_{1}$ is suppressed by a
coupling between torsion and magnetic diffusion constant $\eta$, as
can be seen from the second term of the exponent in the RHS of that
equation. At the $t=0$ these fluctuations are rather suppressed as
one goes along the loop. At the end of the loop where $s=L$ ,
$b_{1}$ is very weak and suffers a strong damping. Analogous result
holds for $b_{2}$ and therefore for $\textbf{B}_{1}$. Despite of the
fact that the initial amplitudes of the fluctuations are great since
$b_{1}\approx{{\tau}_{0}}^{-1}$. Since $v_{0}={\tau}_{0}$ the
Reynolds number Rm is
\begin{equation}
{Rm}=\frac{{\tau}_{0}L}{\eta}\label{40}
\end{equation}
which for a Reynolds Rm number in coronal solar loop regions of
$R_{m}$ and the above torsion yields a diffusion constant of
${\eta}\approx{10^{-8}}$, which is a very small diffusion as happens
in the highly conductive coronal solar region. This shows that the
model presented here is compatible with the fact that the dynamo
begins his action in the convective zone and undergoes magnetic
buoancy to raise to coronal regions of the Sun. Concerning finally
the Zeldovich anti-dynamo theorem one notes that in all above
solutions , the vanishing of torsion implies that turbulence is
planar and the magnetic field cannot be amplified and any dynamo
action is suppressed.

\section{Conclusions} Vishik's idea that the non-stretched dynamos
cannot be fast is tested once more here, by showing that the
fluctuations modes of the solar dynamos are strongly suppressed when
the long wavelength dynamo modes, or small dynamo wave numbers are
effective. Stretching are therefore fundamental in the effective
dynamo convective region of the Sun \cite{6}. The twist or Frenet
torsion is also small, and slow dynamo are shown to be present in
this astrophysical loops. Thus ${\alpha}$-dynamo model is suppressed
by helical turbulent diffusion in dynamo waves. Simplifications in
the model as the constant toroidal flow along the plasma flow
velocity along the tube axis and the small constraint of twist given
by Lopez-Fuentes et al \cite{13} are given. STF Zeldovich-Vainshtein
\cite{14} fast dynamo generation method ,is not the only Riemannian
method that can be applied as in Arnold's cat map but other
conformal fast kinematic dynamo models as the conformal Riemannian
one \cite{15} has been recently obtained. Small scale dynamos in
Riemannian spaces can therefore be very useful for our understanding
of more large scale astrophysical dynamos. Other applications of
plasma filaments such as stretch-twist and fold fractal dynamo
mechanism which are approximated Riemannian metrics have been
recently put forward by Vainshtein et al \cite{14}. Here one has
relaxed the stretch used in STF fast dynamo method and instead use
the NTF kinematic slow dynamos in Frenet frame, showing that small
torsion is fundamental for the suppression of fluctuations. Finally
one has shown that the fluctuations suppression, results frow an
interaction between the magnetic diffusion constant and the Frenet
torsion.
\section{Acknowledgements}
I thank financial supports from Universidade do Estado do Rio de
Janeiro (UERJ) and CNPq (Brazilian Ministry of Science and
Technology).

  \end{document}